\documentclass[english,aps,prper,reprint,showpacs,titlepage,longbibliography]{revtex4-2}

\usepackage[T1]{fontenc}	
\usepackage[latin9]{inputenc}	
\usepackage{geometry}		
\usepackage{graphicx}
\usepackage[above,below]{placeins}	
\usepackage{times}

\usepackage{amsmath}

\usepackage{xcolor}

\usepackage{hyperref}  
\hypersetup{colorlinks=true,urlcolor=blue,citecolor=blue,linkcolor=blue}   
\usepackage{enumitem}          
\setlist{nosep}                 

\begin{document}

\begin{titlepage}

  \title{Grades, equity and graduation rates: a quantitative analysis of the association of learning assistants with improved student outcomes}

  \author{Cassandra A. Paul}
  \affiliation{Department of Physics \& Astronomy | Science Education Program, San Jose State University, One Washington Square, San Jose, CA,
    95192} 
  \author{David J. Webb}
   \affiliation{Department of Physics \& Astronomy, University of California, Davis, One Shields Avenue, Davis, CA, 95616}


  \begin{abstract}
  In this paper we use multilevel modeling of 10 introductory courses from four STEM disciplines to quantify associations of Learning Assistants (LAs) with the student-level outcomes, pass-rates, and retention. Overall we find that having an LA in a class is associated with improvement of these metrics for all students. Furthermore, along with all students showing improvement, we find noticeable demographic differences in the amount of improvement. For instance, LAs seem to be associated with larger improvements in pass-rates for historically marginalized groups, and larger increases in graduation rates for women identifying as belonging to historically marginalized races and ethnicities when compared to students from the same group but with class sections without LAs. This indicates that Learning Assistant programs may be an excellent investment for diverse institutions.
    \clearpage
  \end{abstract}

\maketitle
\end{titlepage}

\section{Introduction}
Over the past two decades, the Learning Assistant (LA) model, in which undergraduate students who have previously succeeded in a course serve as discussion facilitators, has been widely implemented in physics departments and other disciplines nationwide \cite{oteroWhoResponsiblePreparing2006}. The presence of a Learning Assistant in a given classroom has been shown to correlate with improved learning outcomes measured by conceptual inventories \cite{margoniner_2020,VanDusen2016}, grades and DFW rates \cite{alzen_2018, alzenLogisticRegressionInvestigation2018} (including one causal study \cite{kramer_establishing_2023}), and has also been shown to correlate with social aspects of learning like belonging and confidence \cite{clements_impacts_2025}, and other related goals of the program \cite{barrassoScopingReviewLiterature2021}. Emerging research has indicated that Learning Assistants may impact overall retention and graduation rates \cite{tedeschi_improving_2023}, but doubt has been cast on this finding as well \cite{bundyAnalyzingLearningAssistant2025}. In this large quantitative study at one diverse regional public university, we add to this research on possible effects of Learning Assistants on grades and graduation and broaden the results by separately considering different demographic groups. This is an important contribution because studies have shown that research on physics students is disproportionally done at research 1 (R1) institutions on significantly less diverse student populations \cite{Kanim2020}.   

\section{Theoretical Framework}
Recent research \cite{webb_attributing_2023} suggests that differences in achievement between different demographic groups in introductory science courses might be completely attributable to the structures of these courses. Adopting such a Course Deficit Model \cite{Cotner2017}, rather than the common Student Deficit Model, is useful for instructors because changes to the structures of courses are relatively easier to make than changes to the students. Research has already found that changes in assessment \cite{Cotner2017, salehi_gender_2019, singh_test_2021}, course organization \cite{Webb2017,webb_attributing_2023, paul_examining_2025}, and in grading \cite{Paul2022, webb_highstakes_2025, Simmons2020} can decrease (or even collapse) demographic differences in achievement at the same time that they improve achievement for each demographic group and without changing intellectual standards. Importantly, the Course Deficit Model doesn't posit that all students have similar backgrounds and experiences, instead it assumes that all students regardless of demographic background are equally capable of being successful in a given course.  
We use two equity models in this paper. We employ Equity of Individuality \cite{Rodriguez2012} where we evaluate each demographic group's success in classes with LAs compared to those without, and also Equity of Parity \cite{Rodriguez2012} where we compare the success of different demographic groups with the idea that different groups can have equal success under the best course structures.

\section{Research Questions}
The overall point of this first-look at courses which use an LA is to look for correlations between a student seeing an LA in a class and successes in that student's academic career.  We examine two types of success.

1) Is the presence of an LA in a course correlated with enabling students to continue in their academic career without needing to retake their course? 

2) Is the presence of an LA in a course correlated with enabling students to graduate in their STEM major?

\section{Data and Methods}
Learning Assistants were deployed in ten large introductory STEM courses and in fifteen additional non-introductory STEM courses at a large regional public university. In this paper we focus on the ten introductory courses that include eleven years of an introductory chemistry course, ten years of an introductory biology course, thirteen years each of six introductory physics courses, and eight years each of two introductory computer science courses. All together we include 615 lecture sections that did not use LAs anywhere in the course and 67 lecture sections that used LAs somewhere in the course (most often during lecture). The dataset includes 25,098 unique students with 8,883 present only once in the database, 6785 present twice, 4573 appearing three times, 3522 appearing four times, 1023 appearing five times, and the other 312 appearing from 6-9 times.

We obtained data on gender, race/ethnicity, course grades, GPA, and graduation status from the university administration. There are 16,281 men in the database, 8,772 women, 30 nonbinary students, and 15 for whom we have no gender information. We define historically marginalized (HM) racial/ethnic groups as we have previously \cite{paul_examining_2025} and find 6,475 students from the HM groups, African-Americans, Latina(o), Chicana(o) or Mexican American, Hispanic-other, Native American, and Pacific Islander students, and 18,609 from non-marginalized racial/ethnic groups, and 15 for whom we have no data.

These data will be used to estimate WFD (withdrawal, F-grade, or D-grade) rates as well as STEM graduation rates. For both WFD rates and graduation rates we will need to use Multilevel Modeling, rather than ordinary regression or ordinary logistic regression, to properly estimate both the rates and the standard errors.

The WFD rate varies across classes (i.e. lecture sections) so we need to include the class-to-class variations in the error estimates when comparing classes with LAs to classes without LAs. This is accomplished with Multilevel Modeling by grouping the students within classes so that student is the lowest level of the hierarchy and their class is the next higher level and the course is the highest level. Class-level effects are computed first and then averaged over classes within the course and then the overall effects of the 10 courses are computed. We use STATA software which allows us to do all of these steps together in one fit.

To estimate the student graduation rates we will also use multilevel modeling. The 10 courses we are using enroll students from different majors and at somewhat different points in their academic careers. This leads to overall graduation rates that depend on the course, so we need to take this into account. For this reason we group the students (lowest level) into their course section (second level) which are then grouped by course (highest level). So the fitting procedures first fits by class and then by course and then averages over the 10 courses.  In addition, students who transferred to the school have different graduation rates than those admitted as Freshmen. We will leave out transfer students in this first look at graduation rates.

We also note that both WFD and graduation are binary variables (e.g. either student graduated or they didn't) so we will be using a multilevel version of logistic regression to calculate both the odds of graduation (or of WFD) and the change in the odds when students had an LA in their class and use these values and their associated standard errors to calculate the relevant probabilities (i.e. odds of grad = (probability of grad) / (1 - probability of grad)). We will be displaying these probabilities, and/or differences in probabilities, along with their standard errors.

\section{Results}
\subsection{Withdrawals, D-grades, and F-grades}
\begin{figure} [htb]
\includegraphics[trim=3.4cm 3.0cm 5.4cm 3.6cm, clip=true,width=\linewidth]{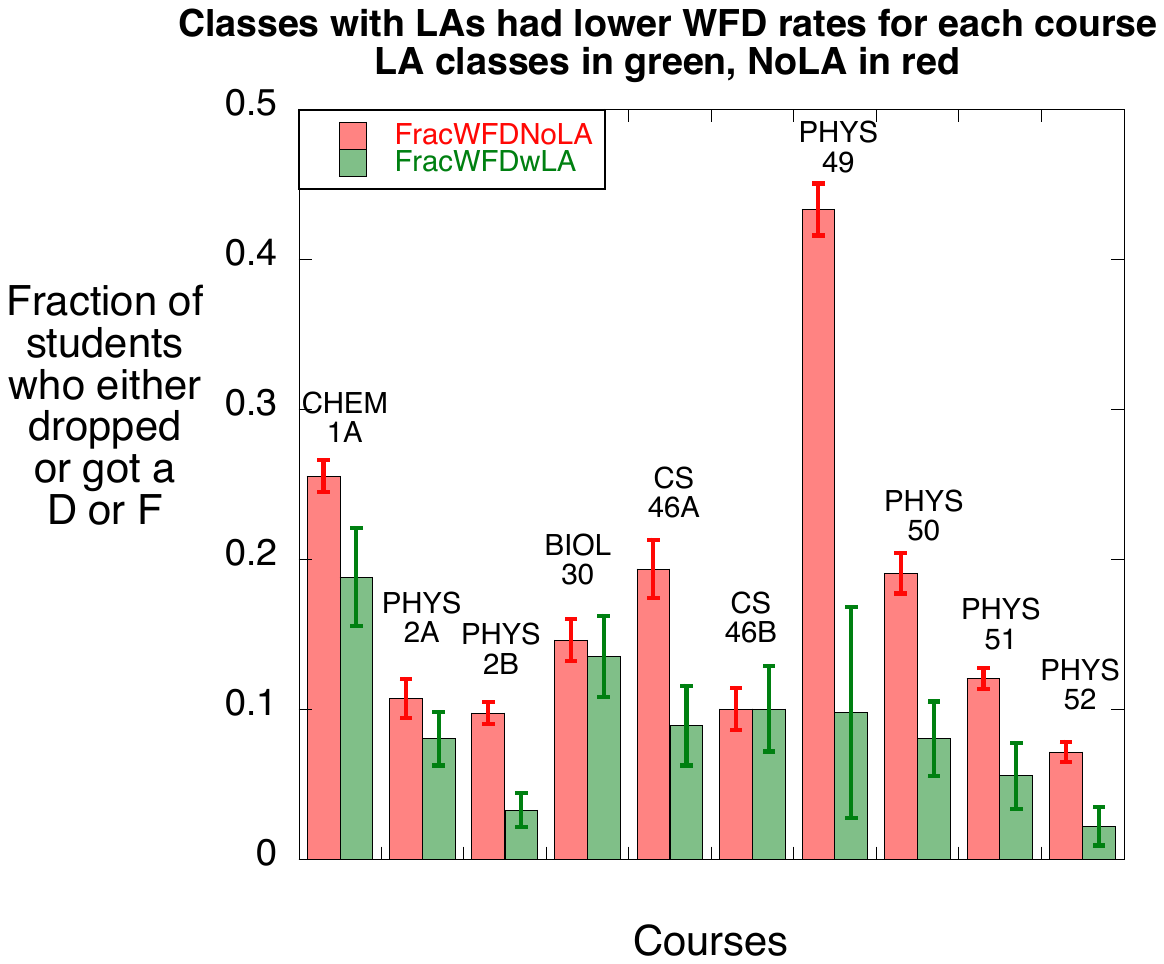}
\caption{The fraction of students in a class who either withdrew from the course or received an F or D grade is shown for each of the 10 introductory courses that used LAs. Classes using LAs (green) and classes that did not use an LA (red) are shown side by side. The error bars are standard errors.}
\label{fig:WFDRates}
\end{figure}

Proceeding smoothly in one's academic career involves neither withdrawing nor being given a low grade in any required course. We can see if the presence of LAs has any correlation with this issue by finding the WFD rates for each of the 10 introductory courses that used LAs.  For these calculations, we use multilevel modeling with only two levels, student and class for each of the 10 courses.  The model we use for each course includes a variable $LA$ = 1 for classes with LAs and = 0 for classes without and looks like 
\begin{equation}
Log(OddsWFD)  = b_0 + b_{LA}LA 
\label{eq.WFDOdds}
\end{equation}
where $e^{b_0}$ gives us the average WFD odds in the non-LA classes and $e^{b_{LA}}$ = (Odds WFD for LA classes) / (Odds WFD for nonLA classes).  We compute the average fraction of students in the WFD group using $OddsWFD = (FractionWFD)/(1-FractionWFD)$ and propagate errors to find the results shown in Figure \ref{fig:WFDRates}.  The figure shows that classes with LAs in them all had lower WFD rates than the same course classes that did not use LAs.  The lower WFD rates between LA and nonLA classes are significant at the $P = 0.05$ level for all of the classes except for Chem 1A, Phys 2A, Biol 30, and CS46B.  We do see that there seems to be large variation in the WFD rates across the courses.  We can't explain the reasons for these differences without qualitative data (such as observations or interviews) but we suspect that some of these differences have to do with how often and in what ways the LAs are interacting with students in the class, as well as features that are unique to the course (i.e. nature of the content, format of the course) the course section (i.e. layout of the classroom, time of the section), or the instructor (pedagogical values and practices, assessment strategies). In Fig. \ref{fig:WFDDemog} (which is discussed in more detail in Section \ref{sec.Demographics}) we compute changes in WFD rates for different demographic groups when an LA is present and we can use those results (for instance add together the two gender results) to draw an overall numerical result. For a (hypothetical) situation where all classes in these 10 courses used LAs compared to a situation where none of the 10 courses used LAs, we estimate that 304 $\pm$ 31 student grades each year would have moved above the WFD level.

\subsection{Graduation rates}
The ultimate aim of most students is graduation in their chosen major.  As we have discussed, there are a couple of complications with these graduation rate calculations.  As noted we are using multilevel modeling to group students at the highest level by the course they are taking and, within the course, the actual class they were enrolled in because it is the class which may or may not have an LA in it.  A second issue is that students may have seen LAs in one class or in more than one class.  In order to make the simplest general estimate of any differences in graduation rates for students who have seen an LA, we will define one LA-dependent variable, $LASomewhere$ = 1 if the student has had an LA in any class they took and = 0 if they have not.  We will also include each student only once in our analyses. For each student appearing more than once we randomly choose which class to use and then test several different random choices to make sure the results don't depend on these random choices. As discussed in the methods section, we will limit our study to include only students who were admitted as Freshmen. With these constraints we ask is there a difference in STEM graduation rates for students who start in a STEM major and have seen an LA in a class vs those who also start in STEM but haven't seen an LA.  To compare graduation rates we use 3-level modeling we have described to fit the following model:
\begin{equation}
Log(GradOdds) = b_0 + b_{LASomewhere}LASomewhere
\label{eqn:GradRates}
\end{equation}
where the coefficient, $b_{LASomewhere}$ from the fit will provide us with the appropriate odds ratio to measure the difference in graduation rates 
\begin{equation}
\begin{split}
\text{LAvsNonLAOddsRatio} & = e^{b_{LASomewhere}} \\ & = \frac{\text{GradOddsWithLA}}{\text{GradOddsNoLA}}
\end{split}
\label{eqn:Ratio}
\end{equation}
An odds ratio $> 1$ tells us that students who have seen an LA ended up graduating at a higher rate.

We use this model for each of four different standardly-defined graduation rates, graduation within 4-years, graduation within 5-years, etc. for 6 and 7 years with the student's admission dates appropriately constrained so that graduation was possible in the number of years we consider. The odds ratio, $e^{b_{LASomewhere}}$, for each of these four graduation rates are shown in Table \ref{tab:GradOddsRatios} along with the estimated standard errors and P-values.  It seems like having an LA in a student's course is correlated with higher graduation rates (odds ratio > 1) for all of these graduation definitions with 5-year and 6-year graduation rate differences being statistically significant at a level higher than 99.9\%.  Using the Eq. \ref{eqn:GradRates} fitted coefficients, the no-LA 6-year STEM graduation probability works out to be 61\% $\pm$ 5\% and the rate when a student has seen an LA is 70\% $\pm$ 5\%.  The errors on these overall rates are dominated by the differences between the graduation rates when averaging over the 10 courses.

\begin{table}[htbp]
\caption{The odds ratios, OddsWithLA/OddsWithoutLA, are shown for each of 4 standard graduation rates, 4-years or less, 5-years or less, etc.  Standard errors and P-values are also shown.  An odds ratio > 1 happens if students who have seen an LA, in one of the 10 courses included in our analyses, graduate at higher rates than students who never saw an LA in one of those courses.}
\label{tab:GradOddsRatios}
\begin{ruledtabular}
\begin{tabular}{c c c c}
\textbf{Rate} & \textbf{OddsRatio} &\textbf{StandardError} & \textbf{P-value}\\ 
\hline
4-yearGrad & 1.27 & 0.11 & 0.006 \\
5-yearGrad & 1.40 & 0.10 & $<10^{-3}$ \\
6-yearGrad & 1.47 & 0.13 & $<10^{-3}$ \\
7-yearGrad & 1.41 & 0.16 & 0.002 \\
\end{tabular}
\end{ruledtabular}
\end{table}


\subsection{\label{sec.Demographics}Demographic differences}
\begin{figure} [htb]
\includegraphics[trim=3.2cm 3.2cm 5.6cm 3.6cm, clip=true,width=\linewidth]{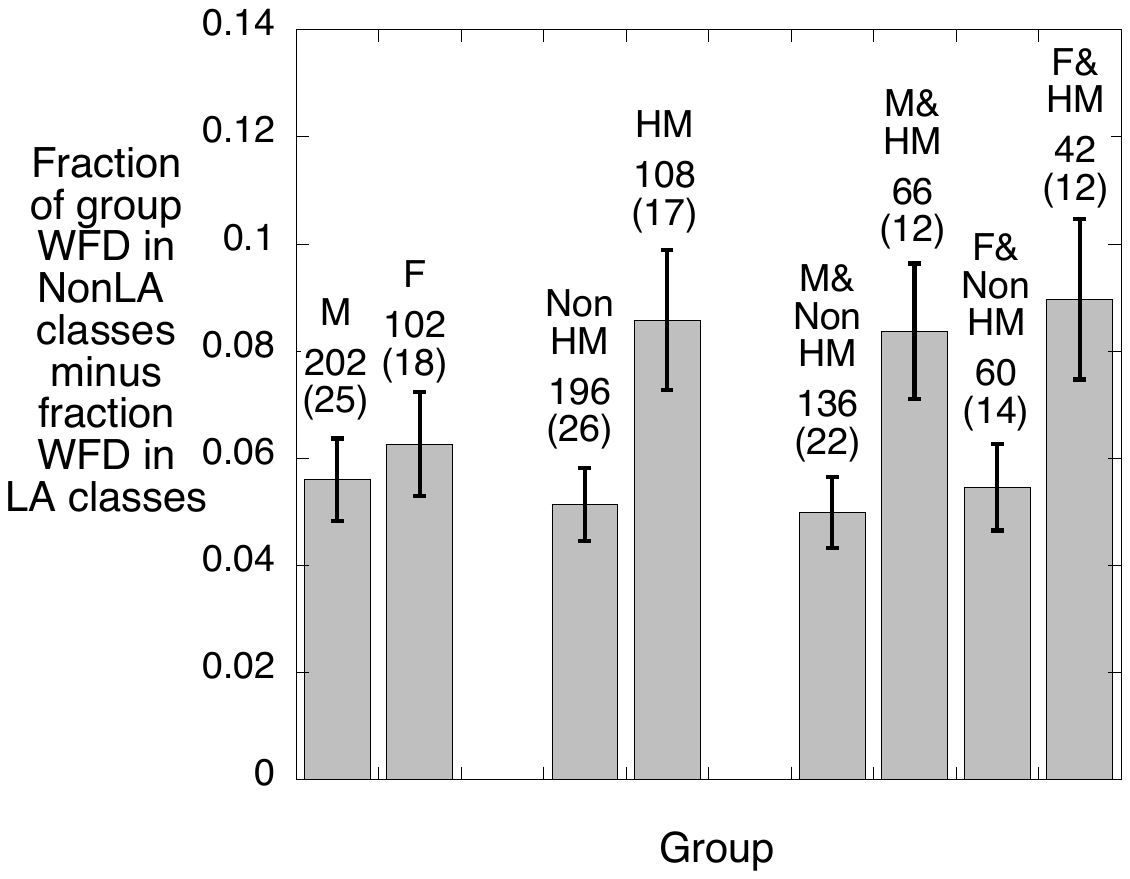}
\caption{The estimated fraction of students who were in the WFD category in classes without LAs but would not have been if they had been in classes that included LAs.  Below each group label we estimate of the number of students per year (with the standard error in parentheses) who would have made up that fraction if LAs were present in \textbf{every} class in each of these ten courses.  The error bars are standard errors.}
\label{fig:WFDDemog}
\end{figure}

Considerable research supports the idea that the way that a course is structured can lead to biases toward some demographic groups (i.e. \cite{webb_attributing_2023, Simmons2020, Cotner2017, salehi_gender_2019, VanDusen2016}). A result of this is that changing the structure of a class, such as including LAs in the teaching process, may benefit some demographic groups more than others. We find examples of this situation in the data we are presenting here.
We can estimate the differences in WFD rates in LA classes compared to non-LA classes for different demographic groups by using Equation \ref{eq.WFDOdds} but include a higher layer in the multilevel modeling by grouping the classes into the 10 courses.  In Fig. \ref{fig:WFDDemog} we show the fraction of students in eight different demographically-defined groups who find themselves in the WFD group in the non-LA courses but seem to be more successful in their course when a class has LA(s). In Fig. \ref{fig:WFDDemog} the groups are labeled by gender (M or F) and by historically marginalized race/ethnicity status (HM or NonHM). The total numbers of additional students per year who might no longer fail by changing from no classes having LA to all classes having LAs is computed on a course by course basis using Eq. \ref{eq.WFDOdds} and then combined to give the numbers quoted in the figure. The gender difference seems too small to measure but the differences in the fractions of students from historically marginalized racial/ethnic groups and their peers is much larger.  For every demographic group listed the WFD difference between classes with LAs and classes without LA (as represented by the positive value of the bar graph) is statistically different from zero with P-values each equal to or smaller than $10^{-3}$.

In the same way that we compared 6-year graduation rates for the whole class, we can also estimate the graduation rates for the different demographic groups.  Again, we consider only students who are in STEM during their introductory classes, and were admitted as Freshmen early enough to have graduated within 6 years by Spring 2025.

\begin{figure} [htb]
\includegraphics[trim=3.2cm 3.0cm 5.6cm 4cm, clip=true,width=\linewidth]{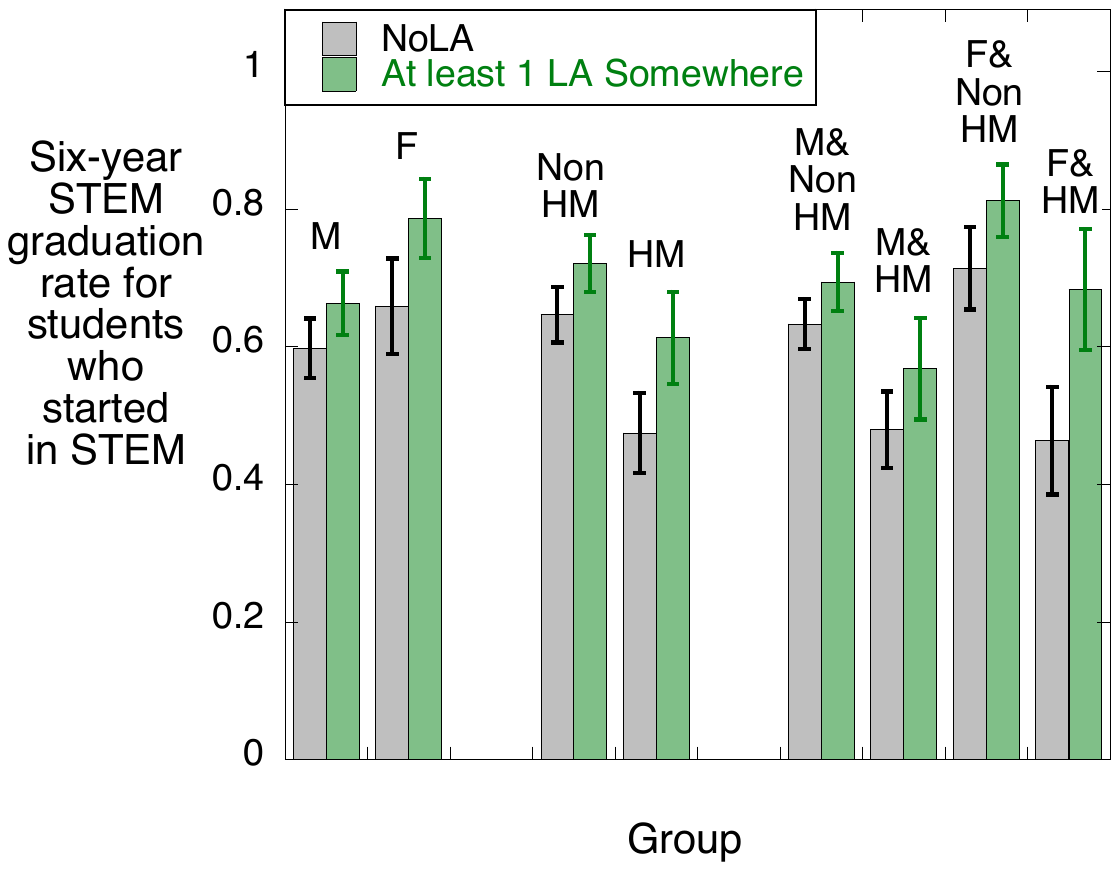}
\caption{Estimated 6-year graduation fractions for the same demographic groups as shown in Fig. \ref{fig:WFDDemog}.  The error bars are standard errors and are dominated by the averages over the base (no LA) probabilities of the 10 different courses.}
\label{fig:STEMGradDemog}
\end{figure}

The rates for graduation in a STEM major are shown in Fig. \ref{fig:STEMGradDemog} for the same eight demographic groups we used for the WFD figure.  The largest increases are seen for students who are members of racial/ethnic group historically marginalized in STEM. 

\section{Discussion}
Our finding that the presence of a Learning Assistant in an introductory course is correlated with increased course pass rates and graduation rates implies that Learning Assistants are a factor in increasing student success in STEM, especially for historically marginalized students. Because each demographic group improves both in pass-rates, and in graduation rates, the LA program achieves equity of individuality. While there is not yet compelling evidence of achieving equity of parity for all demographic groups, preliminary evidence suggests that the LA program is correlated with decreasing graduation equity gaps, especially for women from historically marginalized groups. This information is useful for administrators, especially those with diverse student populations, looking to simultaneously increase overall student success as well as equity. Interestingly, our data show that the LA program seems to correlate with larger effects for students from historically marginalized groups when considering WFD rates, but instead correlates with larger effects for women, especially those from marginalized groups when considering graduation rates.  
Our data suggest that putting Learning Assistants in all introductory courses could decrease course level fail rates by up to 300 students per year in a program that is currently limited to a single college at a regional R2 institution. This is likely an upper limit on LA benefits as not all instructors want an LA, and some course formats are not conducive to LA support.

\section{Limitations}
Despite the potential success of the LA program, there are several limitations to this work. For one, it is done at a single institution, so we can only generalize in the sense that we assume that the LA program here is similar to national programs. The period of study also  includes a pandemic, when LAs worked in much more unconventional ways. Finally, the study does not account for differences between instructors, course structures, or time spent with LAs to name just a few of the numerous classroom characteristics that might also impact student retention. There are quite possibly selection effects in these data that we have yet to identify. Future research will target some of these class features in order to consider their overall correlations with student retention.

\section{Conclusions}
We find that sections with LAs generally have lower fail-rates than those without. We see an increase in 6-year STEM graduation rates by about 9 percentage points for students who have had an LA in one of their classes vs. those who haven't. Finally, we find that the effect of having an LA in a course section is larger for students from historically marginalized groups, women, and especially students who are both women and from historically marginalized groups for each of the two metrics we measured. For example, women from historically marginalized groups who have an LA in at least one of their courses increase their chances of graduating with a STEM degree by about 20 percentage points. Future research will investigate possible alternative effects discussed in our limitations section.

\acknowledgments{The authors would like to thank Dr. Ashley Defensor for her work in tracking LA supported courses, and the San Jose State University Physics Education Research group and the conference reviewers for feedback on this work. This material is based upon work supported by the National Science Foundation under Grant No. 2234071 and the now terminated Grant No. 1953760. This work would not have been possible without the support of San Jose State University's Research, Scholarship, and Creative Activity Assigned Time Program and the Research and Innovation Scholarly Entrepreneurship (RAISE) award, which allowed us to continue investigations on historically marginalized students after National Science Foundation Funding was pulled for these efforts.}









\bibliography{LABibFile} 

\end{document}